\documentclass[graybox]{svmult}

\usepackage{type1cm}        
\usepackage{algorithm}
\usepackage{algpseudocode}
\usepackage{adjustbox}
\usepackage{xcolor}
\usepackage{amsmath}
\usepackage{amssymb}
\usepackage{multirow}

\usepackage{comment}
\usepackage{xcolor}
\usepackage{makeidx}         
\usepackage{graphicx}        
\usepackage{caption}                            
\usepackage{subcaption}
\usepackage{multicol}        
\usepackage[bottom]{footmisc}

\usepackage{newtxtext}       %
\usepackage[varvw]{newtxmath}       
\usepackage[numbers]{natbib}
\newcommand*{\affaddr}[1]{#1} 
\newcommand*{\affmark}[1][*]{\textsuperscript{#1}}

\makeindex             

\begin{document}

\title*{A Transformer-based Deep Learning Algorithm to Auto-record Undocumented Clinical One-Lung Ventilation Events}
\author{Zhihua Li \protect\affmark[1], Alexander Nagrebetsky \protect\affmark[2], Sylvia Ranjeva \protect\affmark[2], Nan Bi \protect\affmark[1], Dianbo Liu \protect\affmark[3], \\ Marcos F. Vidal Melo \protect\affmark[4], Timothy Houle \protect\affmark[2], Lijun Yin \protect\affmark[1], Hao Deng \protect\affmark[2] \newline
\affaddr{\affmark[1]\small Binghamton University}
\newline
\affaddr{\affmark[2]\small Massachusetts General Hospital}
\newline
\affaddr{\affmark[3]\small Mila AI Institute}
\newline
\affaddr{\affmark[4]\small Columbia University Irving Medical Center}
\newline
}


\institute{
\email{zli191@binghamton.edu;
anagrebetsky@mgh.harvard.edu;
sranjeva@mgh.harvard.edu;
nbi1@binghamton.edu;
dianbo.liu@mila.quebec;
mv2869@cumc.columbia.edu;
thoule1@mgh.harvard.edu;
lijun@cs.binghamton.edu;
hdeng1@mgh.harvard.edu
}}

%
%
\authorrunning{Zhihua Li et al.}
\titlerunning{One-lung Ventilation Event Detection Using Transformers}
\maketitle
\vspace{8mm}
\abstract*{Each chapter should be preceded by an abstract (no more than 200 words) that summarizes the content. The abstract will appear \textit{online} at \url{www.SpringerLink.com} and be available with unrestricted access. This allows unregistered users to read the abstract as a teaser for the complete chapter.
Please use the 'starred' version of the \texttt{abstract} command for typesetting the text of the online abstracts (cf. source file of this chapter template \texttt{abstract}) and include them with the source files of your manuscript. Use the plain \texttt{abstract} command if the abstract is also to appear in the printed version of the book.}
\vspace{-16mm}
\abstract{As a team studying the predictors of complications after lung surgery, we have encountered high missingness of data on one-lung ventilation (OLV) start and end times due to high clinical workload and cognitive overload during surgery. Such missing data limit the precision and clinical applicability of our findings. We hypothesized that available intraoperative mechanical ventilation and physiological time-series data combined with other clinical events could be used to accurately predict missing start and end times of OLV. Such a predictive model can recover existing miss-documented records and relieves the documentation burden by deploying it in clinical settings. To this end, we develop a deep learning model to predict the occurrence and timing of OLV based on routinely collected intraoperative data. Our approach combines the variables' spatial and frequency domain features, using Transformer encoders to model the temporal evolution and convolutional neural network to abstract frequency-of-interest from wavelet spectrum images. The performance of the proposed method is evaluated on a benchmark dataset curated from Massachusetts General Hospital (MGH) and Brigham and Women's Hospital (BWH). Experiments show our approach outperforms baseline methods significantly and produces a satisfactory accuracy for clinical use.}
\begin{figure*}[t!]
	\centering
		\includegraphics[scale=0.34]{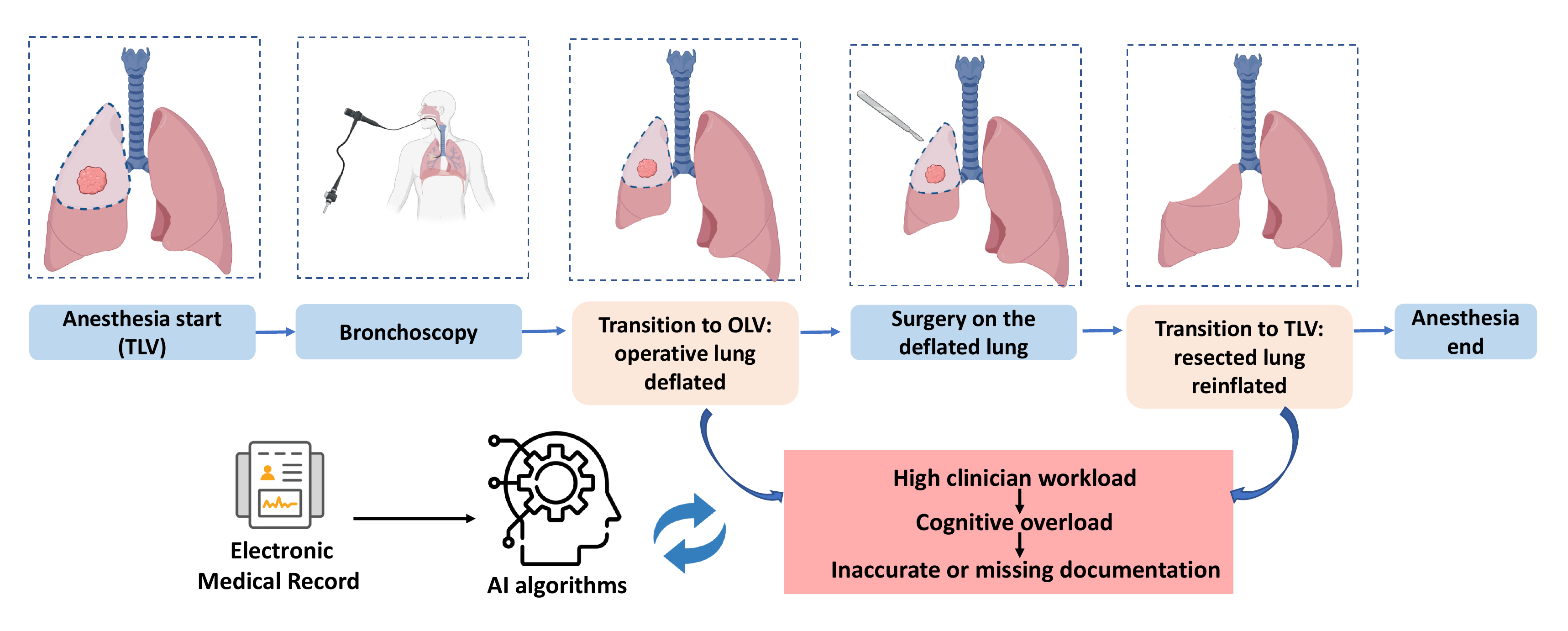}
	  \caption{Intraoperative sequence of events during lung resection with one-lung ventilation. TLV: two-lung ventilation; OLV: one-lung ventilation.
}\label{fig1}
\end{figure*}
\section{Introduction}\label{intro}
Among two million people diagnosed with lung cancer each year \cite{bray2018global}, approximately one-third need lung resection surgery \cite{perera2021evidence}. An operation on the lung requires one-lung ventilation (OLV) to deflate and immobilize the operative lung for surgical visualization. OLV, in turn, presents unique challenges for mechanical ventilation and for prevention of postoperative pulmonary complications \cite{marseu2016peri}. Transition to OLV during thoracic surgery is a distinct risk factor for post-operative acute lung injury, ranging in severity from mild atelectasis to severe acute respiratory distress syndrome (ARDS) \cite{lohser2008evidence}. Strategies for lung-protective management during two-lung ventilation have evolved from studies of ARDS in ICU populations \cite{allison2015high,neto2016association}. These protective ventilation strategies aim to provide sufficient oxygenation while minimizing ventilator-induced alveolar trauma, inflammation, and cyclic collapse \cite{slutsky2013ventilator}. While lung-protective strategies for two-lung ventilation are well-described, patients receiving OLV during lung resection, a cohort that is inherently vulnerable to pulmonary complications, suffer from a paucity of clinically meaningful evidence.

The lack of reliable data on the time of transition from two-lung ventilation to OLV is a limiting factor that complicates research into the pathophysiology and prevention of pulmonary complications after lung surgery.
The transition from two- to one-lung ventilation is a time of heightened risk for respiratory decompensation, and is thus a time of high cognitive and procedural burden for the anesthesia provider. Therefore, the manually entered documentation of this transition may not be timed correctly and is often missing (as shown in Fig \ref{fig1}). At the same time, multiple streams of physiological data that are recorded automatically during the start and end of OLV, make it possible to accurately impute the occurrence and timing of OLV. For example, airway pressures, volumes of delivered breaths, and respiratory rate change within seconds after the start and end of OLV. Other physiological metrics such as heart rate, blood pressure, exhaled CO2, and hemoglobin oxygen saturation (SpO2) may also change in response to the start and end of OLV.
The missing or incorrectly timed documentation of OLV illustrates a common clinical scenario where the need for event documentation in the procedural or emergency care settings competes for clinicians' attention with patient care tasks. Furthermore, the need to recall and document clinical events contributes to the cognitive overload of clinicians and can result in burnout
\cite{iskander2019burnout}. The cognitive burden of clinical documentation may be alleviated by a machine learning algorithm that automatically detects the occurrence and timing of clinical events of interest. If used in real-time or near-real time, such an algorithm can redirect a clinician's attention toward patient care. In retrospective analysis of data, it can aid in clinical quality control and imputing missing data for research.

Given large-scale documented historic medical records, we hypothesized that a data-driven machine learning model could estimate the occurrence and timing of clinical events using routinely available physiological settings and measurement data. We aimed to test this hypothesis by developing a deep learning model and testing its ability to detect the occurrence and timing of OLV start and end in a dataset of patients undergoing lung resection surgery. Specifically, the OLV timestamp prediction can be formulated as a time-series detection task, which takes multiple sequences of 1-D time-series signals as the inputs and outputs the target timestamp of the occurring event. Recognizing a specific segment of the waveform is the key to locating the desired timestamps. 

This paper proposes a Transformer-based deep learning framework for OLV timestamp detection. Apart from spatial features, the 1-D signals are transformed into wavelet spectrum images and fed into customized convolutional neural networks to extract the discriminative frequency information. Furthermore, an innovative label temporal smoothing technique is proposed to optimize the procedure to locate the maximum scores from the prediction curves. Experimental results show that our proposed method significantly outperforms the basic time-series change-point detection methods and the recurrent-based deep learning methods.


\section{Related Works}\label{related}
The objective of this work is to detect if there is an OLV event occurring for every minute of the signals. In the area of time-series analysis, Twitter \cite{vallis2014novel} employed statistical learning to detect anomalies in both applications (Tweets Per Sec) as well as system metrics (CPU utilization). \cite{siffer2017anomaly} proposed an approach to detect outliers in streaming univariate time series based on Extreme Value Theory that did not require hand-set thresholds. \cite{facebook_kat} developed a time-series anomaly detection toolkit by packaging a series of statistic-based methods such as CUSUM (cumulative sum) and Bayesian Online Change Point Detection. However, they are not suitable for the OLV detection tasks because there are no observable change points near the OLV actions for most of the variables. Additionally, some signals have many change points that are not related to the OLV procedure; thus, many false positives would be generated.

Traditional hand-crafted features are expensive to create and require expert knowledge of the field. The performances of traditional statistical models are not satisfactory in real applications. Recently, deep learning approaches have shown superior power in big data analysis with successful applications to computer vision, pattern recognition, and natural language processing \cite{liu2017survey}. Researchers are investigating data-driven models to improve anomaly detection accuracy \cite{gamboa2017deep}. 

Opprentice \cite{liu2015opprentice} used existing detectors to extract anomaly features and fed them to a random forest classifier to automatically select the appropriate detector-parameter combinations and the thresholds.
\cite{xu2018unsupervised} proposed Donut, an unsupervised anomaly detection algorithm based on Variational Auto-Encoder (VAE), making it the first generative-based anomaly detection algorithm. The reconstruction probability was used as an anomaly indicator.
\cite{deldari2021time} employed a contrastive learning strategy for change point detection by learning an embedded representation through self-supervision. \cite{wang2016research} proposed an autoencoder-based deep learning network to learn physiological features and use multivariate Gaussian distribution anomaly detection method to detect anomaly data.
Similarly, \cite{Malhotra2016} proposed a LSTM-based encoder-decoder network to construct a predicted multivariate "normal" time series and used the reconstruction error for prediction.

The existing methods are mostly unsupervised and based on statistical stationarity assumptions. Therefore, they failed to handle the more complicated scenarios for OLV timestamp estimation and capture the correlations between different variables addressed in this paper.
Another barrier using existing methods is that they are incapable of using multiple input signals in a complementary way. The widely used ensemble method that combines predictions of each signal is ineffective since some variables themselves are not discriminative enough for prediction, and they can only provide complementary information for other dominant variables.
\begin{table*}[ht!]
\resizebox{\textwidth}{!}{%
\begin{tabular}{c|c|l}
\hline
\multirow{3}{*}{\textbf{\begin{tabular}[c]{@{}c@{}}Ventilator \\ setting \\ variables\end{tabular}}} & Tidal volume (VTset) & a mechanical ventilator setting which determines the volume goal of each ventilator-delivered breath \\ \cline{2-3} 
 & Respiratory rate (RRset) & a mechanical ventilator setting which determines the number of ventilator-delivered breaths per minute \\ \cline{2-3} 
 & Positive end-expiratory pressure (PEEP) & \begin{tabular}[c]{@{}l@{}}a mechanical ventilator setting which determines the lowest airway pressure after each ventilator-delivered \\ breath\end{tabular} \\ \hline
\multirow{4}{*}{\textbf{\begin{tabular}[c]{@{}c@{}}Respiratory \\ physiology \\ variables\end{tabular}}} & Exhaled tidal volume (exhaledVT) & the measured gas volume returning from the patient’s lungs to the mechanical ventilator after each breath \\ \cline{2-3} 
 & Respiratory rate (RR) & \begin{tabular}[c]{@{}l@{}}measured number of ventilator-delivered breaths per minute, based on the cyclical variations in composition\\  of gas mixture returning to mechanical ventilator\end{tabular} \\ \cline{2-3} 
 & Peak inspiratory pressure (PIP) & maximum pressure in the airway during each breath \\ \cline{2-3} 
 & Hemoglobin oxygen saturation (SpO2) & \begin{tabular}[c]{@{}l@{}}the measured \% of hemoglobin that is saturated with oxygen which indicated how well oxygen is delivered \\ to blood from the lungs\end{tabular} \\ \hline
\end{tabular}%
}
\caption{Variable descriptions. The variables are categorized into ventilator setting variables and physiology measurement variables.}
\label{tab:1}
\end{table*}
To handle the above challenges, we design an innovative Transformer-based model that absorbs multivariate time-series signals and predicts the timestamps of an OLV event under supervision. The proposed method enables direct communications from signal to signal; it also provides direct temporal communications in a self-attention manner by introducing Transformer encoders. 
\section{Data Management}\label{method}
\textbf{Study Design:}
Our study followed a retrospective cohort design. The Mass General Brigham (MGB) Institutional review board (IRB) committee had reviewed the research protocol and exempted the requirement of individual informed consent due to consideration of feasibility and minimal risks to study human subjects (Protocol ID: \#2021P002173). 

\textbf{Inclusion and exclusion criteria:}
{Inclusion criteria:}
(1) patients who were 18 years or older at the time of surgery;
(2) lung resection with one lung ventilation; 
(3) admitted to study site on or after June 15th 2016 and discharged prior to or on June 15th 2021;
{Exclusion criteria:}
(1) age less than 18 years old;
(2) pregnancy;
(3) intraoperative death.
(4) multiple hospital encounters:
(5) multiple OLV-included surgical procedures performed within encounter:
(6) multiple OLV episodes in surgery:
(7) no OLV timestamp data.

\begin{figure}[ht!]
	\centering
		\includegraphics[scale=0.86]{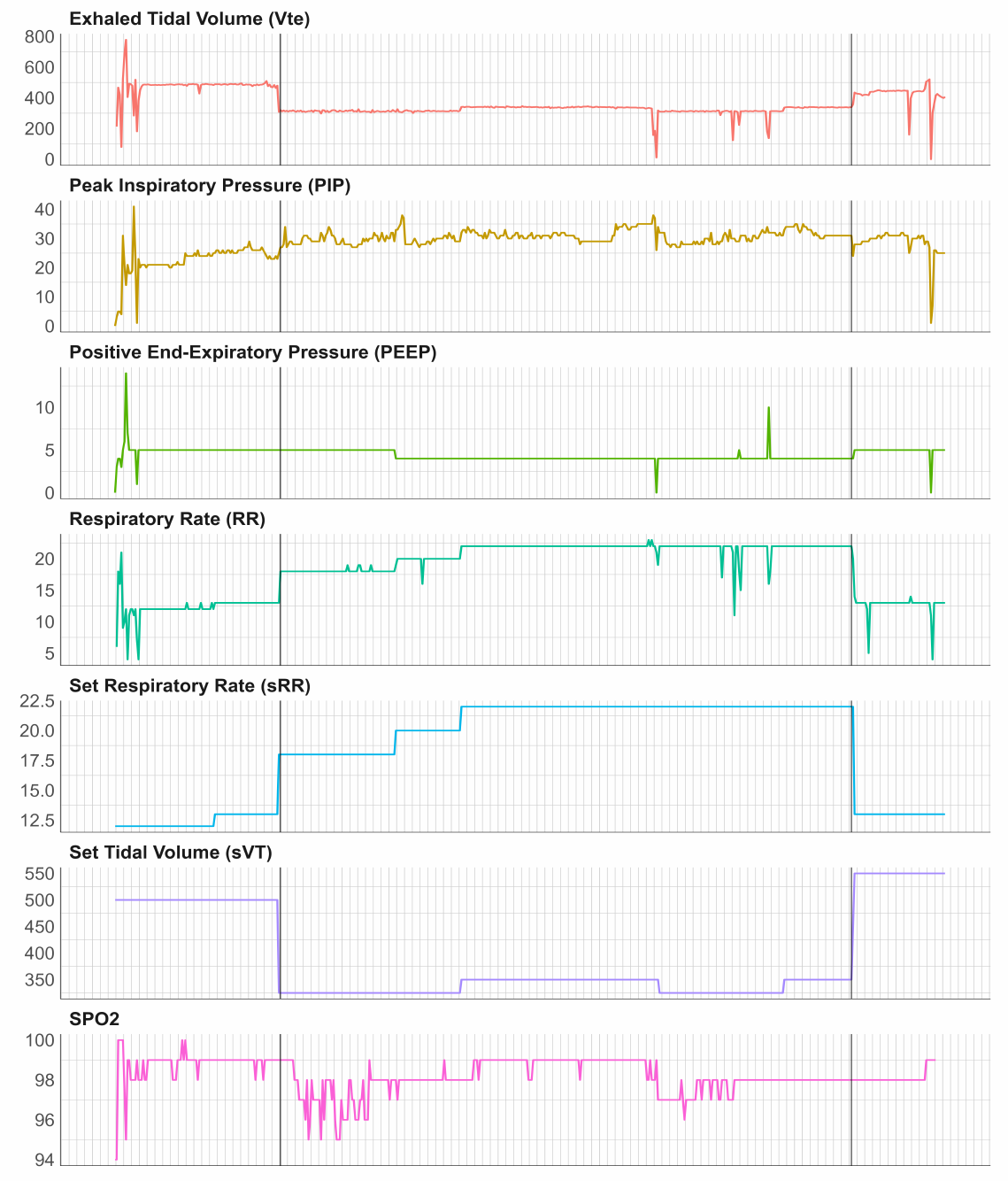}
    \caption{Demonstration of one random patient\'s data record. The first vertical line is OLV start timestamp; the second is OLV end timestamp. The signal sampling rate is 1 sample/min. Each cell marks a five minutes duration.
}
\label{fig2}
\end{figure}

\textbf{Data Source:} Our study team retrospectively extracted Electronic Medical Records (EMR) information from the MGB central data repository named Enterprise Data Warehouse (EDW) utilizing structured SQL queries. We then constructed and maintained a clinical and observational database of all adult patients who received open thoracic surgeries at both Brigham and Women's Hospital (BWH) and Massachusetts General Hospital (MGH) from 2016 to 2021 for this research project. There were no a priori power analyses performed to determine the required sample size due to the innovative model structure. We made all extracted EMR records available for analyses, model development, and validation. The curated dataset consists of 4245 patient admission records.

\textbf{Predictors and Features:}
OLV can lead to immediate changes of value in certain physiological measures such as gas exchange, ventilation mechanics, and hemodynamics. Therefore, modeling the patterns of their changes can be utilized to predict the OLV event. Specifically, we consider variables/physiological measurements that could show an indication of OLV status from the signal shape. We divide the variables into two groups setting values and physiological measurements from biomedical sensors. The former includes (1) VT (tidal volume) set: target volume for each breath, (2) respiration set. The latter contains (1) peak airway pressure: maximum pressure during a breath, (2) measured respiration rate, (3) spo2, and (4) VT exhaled. Detailed variable descriptions are shown in Table \ref{tab:1}.

\textbf{Missing Data:}
Our study collected multi-subject multivariate time-series data, and missingness could occur at any feature throughout the entire observational period of surgical procedures. We assumed that the missing mechanism was under a common assumption of Missing At Random (MAR), and imputed the within-sequence missing values for each feature using the classic linear interpolation method. We used the nearest value padding method to fill in the missing head and tail values of different feature sequences to achieve the same fixed length of time series data for followed modeling steps.

\begin{figure*}[ht!]
	\centering
		\includegraphics[scale=0.34]{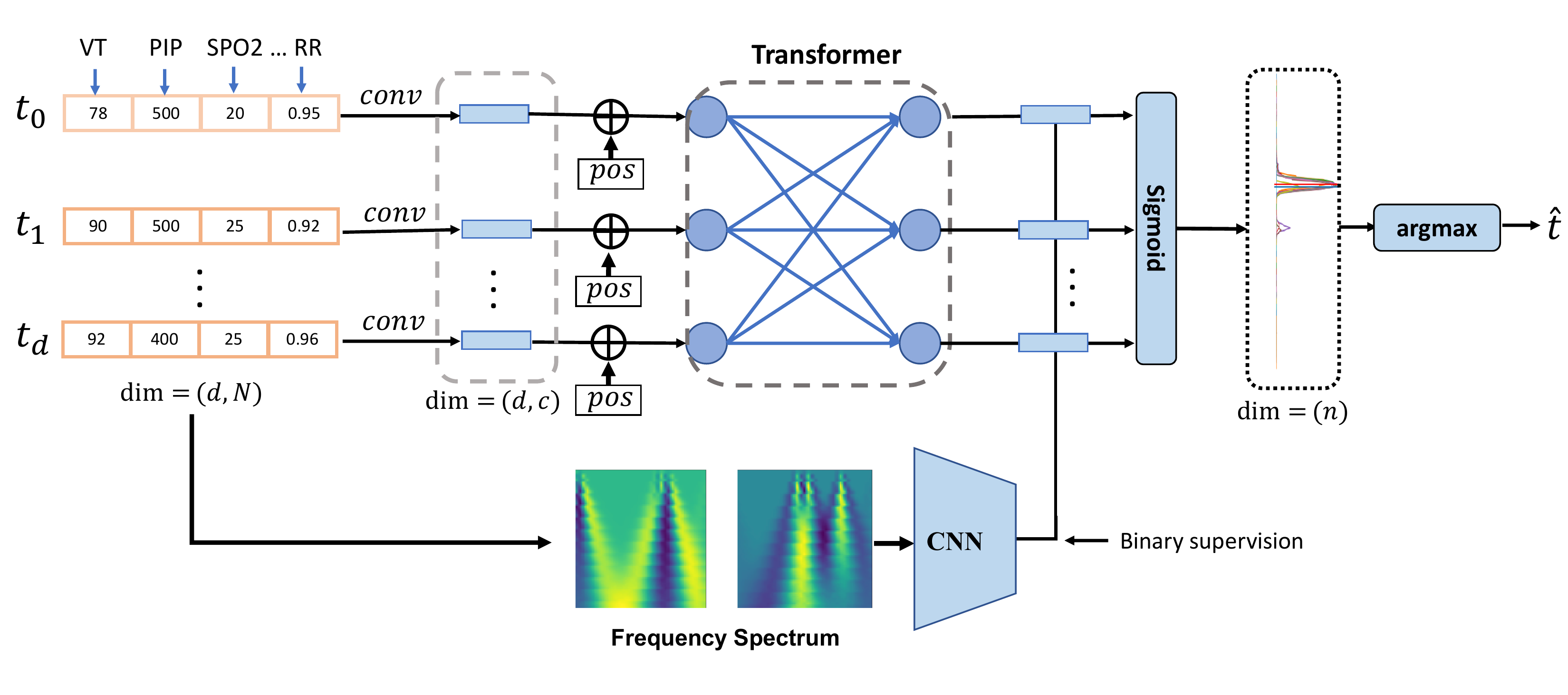}
	  \caption{Framework overview. The upper branch leverages signal morphological characteristics, and the lower branch exploits the frequency spectrum from wavelet transform. First, the multivariate signals are transformed into feature embeddings and fed into the transformer block. We use the combined features from spatial and frequency for the last layer detection. A label smoothing technique is deployed to generate the supervision signals. The timestamp corresponding to the highest OLV occurrence probability is selected as the predicted timestamp. Detailed training and testing steps are shown in Algorithm \ref{alg:cap}.
   }\label{overview}
\end{figure*}

\section{Method}
At a high level, our deep learning model consists of two spatial and frequency branches. The former is primary, and the latter acts as a supplementary. 
For the spatial feature extraction, our method first transforms the variable intensity of each minute to high dimensional feature embeddings, which improves the representation capability. Furthermore, several transformer encoder blocks are stacked to enable spatial-temporal fusion, where the inter-signal relationships are also exploited. In terms of the frequency branch, we apply the wavelet transform to each signal and obtain 2D spectrum images.
In addition, a Gaussian smoothing operation is performed on the ground-truths in order to produce more continuous predictions.
\subsection{Problem definition}
The goal of the OLV documentation system is to automatically predict the start and end timestamp of OLV during lung surgery. Formally, the dataset $P$ with $M$ number of patients is represented by $P = \{p_1, p_2,..., p_M \}$. Each patient $p_{m}$ consists of $N$ variables, and $p_{m} = \{s_0, s_1, s_2,...,s_{N-1}\}$,
where $s_i\in\mathbb{R}^{d}$, and $d$ is the sequence length (in minute) after pre-processing. This paper proposes a deep learning method that attempts to learn a function $\phi(\cdot)$ that absorbs $p$ and predicts the OLV start and end timestamp $y_{s}$ and $y_{e}$, respectively, which can be described as $\widehat{y}=\phi(p)$, where $\widehat{y}$ is an integer and $0\leq y \leq d$.


\subsection{Variable Embeddings}
In practice, the operation time (from surgery start time to surgery end time) varies significantly from patient to patient, causing the patient-wise signal length to unfixed. Feeding an arbitrary length of data to machine learning models needs additional padding steps and brings about more complexity in model building. 
Therefore, we propose to segment the entire sequence into fixed-length windows of length $l_{ws}$, namely a sliding window method with a certain step size $l_{step}$. Specifically, for each training iteration, we randomly sample a start timestamp $t$ where $t<l-l_{ws}$, then select the values from time $t$ to $t+l_{ws}$ as the input to the deep learning model. Since there exist multiple synchronized signal recordings for each patient, we concatenate all the $N$ signals into a channel dimension, resulting in an input $\mathbf{x} \in \mathbb{R}^{N\times l_{ws}}$.

In general, among the $N$ recorded physiological signals, some signals (e.g., tidal volume) provide more clues to address the OLV events than others, while it also happens that informative signals are noisy and the other signals could provide complementary information for prediction. Hence, we decide to represent each signal in a multidimensional way in order to improve the representation capability and capture more complex underlying relationships and signal characteristics. 
To this end,
we employ three linear layers to transform the raw signal recordings $\mathbf{x}$ into high dimensional embeddings, and the embedding dimension gradually increases from $N$ to $64$, and then to $512$, and another $512$ to $512$. 
After encoding the low-dimension raw signal intensities to high-dimension embeddings $\mathbf{f}\in \mathbb{R}^{512\times l_{ws}}$, the densely distributed representation can better represent the signal patterns. Next, the feature embeddings $\mathbf{f}$ are ready for sequential information extraction.

\subsection{Transformer Blocks for Temporal Learning}
To capture the temporal patterns existing around the OLV events, introducing a sequential machine learning model is demanded. Although the recurrent neural networks (RNNs) such as long short-term memory (LSTM) \cite{lipton2015critical} could handle the sequential input, the gating and recurrent steps have shortcomings in modeling long-distance dependencies and could get trapped into gradient vanishing problems. Instead, we deploy a self-attention-based transformer encoder framework. It enables direct interactions among each value in a sequence, thus overcoming the limitations of RNNs and their variants.

The Transformer encoder \cite{vaswani2017attention,devlin2018bert} is composed of multiple identical sub-blocks, and each block consists of a multi-head self-attention module and a fully connected feed-forward layer. Each sublayer is also succeeded by a normalization layer and residual layer. 
Since the transformer blocks are not aware of the input order, we add a learnable positional embedding to the input similar to BERT \cite{devlin2018bert}. And the last layer output from the network can be described as:
\begin{equation}
    \mathbf{f} = \operatorname{MLP}\left(\operatorname{Transformer}\left(\mathbf{f}+\mathbf{x}_{pos} \right)\right)
    \label{trans}
\end{equation}
where $\mathbf{x}_{pos}\in \mathbb{R}^{l_{ws}\times 512}$ is the positional embedding. The final features are obtained by applying a fully connected layer to the output of the Transformer.
\begin{figure}[t!]
	\centering
		\includegraphics[scale=0.3]{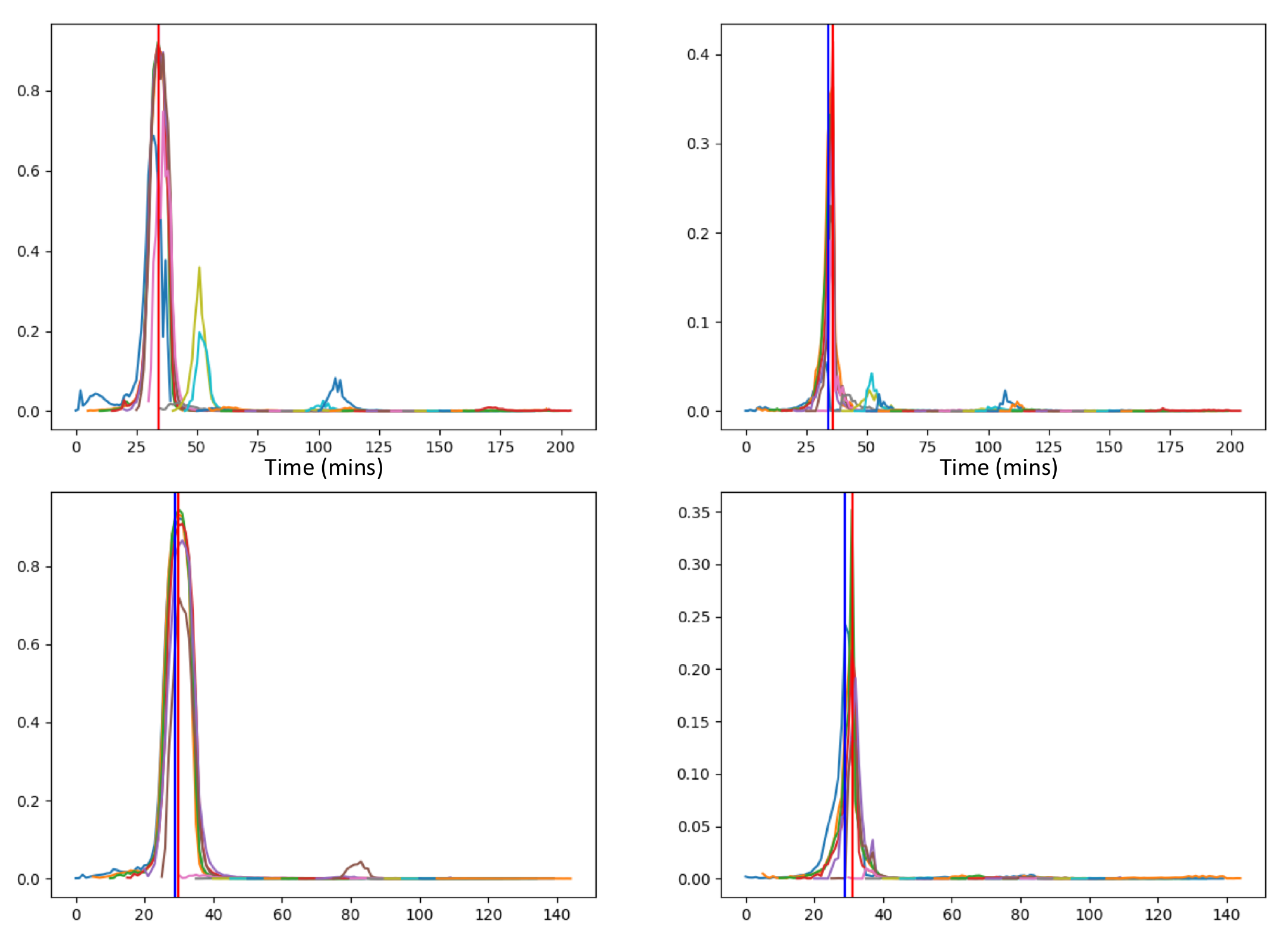}
	  \caption{Scores predicted by the model with (left column) and without (right column) label temporal smoothing. Different color of the curves denotes a different sliding window. Vertical blue line: ground truth timestamps; Vertical red lines: predicted timestamps by finding the maximum .}\label{fig4}
\end{figure}

\subsection{Frequency Domain Features}
In the previous sections, the methods we present are mainly for spatial feature extraction, where we focus on morphological patterns. On the other hand, the change in the spatial domain, namely the frequency domain feature, is a more subject-invariant cue for time-series event detection. 
For example, the power spectrum is a solid feature representation for electrocardiogram (ECG) and electroencephalography (EEG) \cite{banerjee2013application,alquran2019ecg} for classification. 
Since the temporal resolution is crucial for timestamp prediction, we adopt the wavelet spectrum \cite{meyer1992wavelets} 2D image of the input variables as the initial frequency data. By concatenating the $N$ wavelet spectrum images, which are calculated from $N$ input signals, we obtain multi-channel 2D feature maps of size $\mathbb{R}^{100\times100}$. Then, it's straightforward to utilize convolutional neural network (CNN) to extract prediction-related information from the 2D spectrum image. Specifically, we deploy three layers of 2D convolutions together with BatchNorm and non-linear ReLU operations. 
The abstracted frequency features $\mathbf{f}_{freq}\in\mathbb{R}^{64}$ are further concatenated with the spatial features $f$ in Equation \ref{trans} to detect the presence of an OLV event. We describe the calculation as:
\begin{equation}
        \hat{y} = \operatorname{\sigma}\left(\operatorname{CLS}\left(\mathbf{f};\mathbf{f}_{freq} \right)\right)
\end{equation}
where $\sigma$ denotes the Sigmoid operation, and $;$ is feature concatenation. The output logits $\hat{y}\in\mathbb{R}^{l_{ws}}$ are obtained by applying a fully connected classification layer (CLS).

\begin{algorithm}[t!]
\caption{Pseudo code for the proposed model}\label{alg:cap}
\textbf{Training stage:}
\begin{algorithmic}[1]
\State data\_loader samples batches of training records
\For{variables in batch}
\State $t \gets random()$
\State $\textbf{x}_{signal}$ $\gets$ variables[$t$ : $t+l_{ws}$]
\State $\textbf{x}_{freq}$ $\gets$ $\operatorname{wavelet}(\textbf{x}_{signal})$
\State \textbf{f} $\gets$ $\operatorname{Transformer}$($\textbf{x}_{signal}$)
\State $\textbf{f}_{freq}$ $\gets$ $\operatorname{CNN}$($\textbf{x}_{freq}$)
\State $\hat{y} \gets \operatorname{MLP}(f; f_{greq})$ 
\State init\_label = $\operatorname{zeros}$[0 : ws]
\If{$t$ $<$ {$t_{OLV}$} $<$ $t+l_{ws}$}
\State init\_label[$t_{OLV}-t$] $\gets$ 1
\State init\_label $\gets$ Gaussian(init\_label)
\EndIf
\State Update the model by $\operatorname{cross\_entropy}$(init\_label, $\hat{y}$)
\EndFor
\end{algorithmic}

\textbf{Testing stage:}
\begin{algorithmic}[1]
\State data\_loader samples batches of testing records
\For{variables in batch}
\For{$i=1$ to $n_{seg}$}\Comment{calculated by Equation 4}
\State $\textbf{x}_{signal}$ $\gets$ variables[$i$ : $i \times ws$]
\State Repeat step 4 $\sim$ 8 in the training stage. 
\State Obtain $\hat{y}_{seg}^{i}$
\EndFor
\State $\hat{y}=\operatorname{argmax}([\hat{y}_{seg}^{0};\hat{y}_{seg}^{1},...,\hat{y}_{seg}^{n_{seg}-1}])$
\EndFor
\end{algorithmic}
\end{algorithm}

\subsection{Label Smoothing}
Recall that after the entire sequence is split into segments, we need to prepare the ground-truths specifically for each segment. 
First, if the OLV start timestamp $y_{start}$ occurs inside the sampled $t$ to $t+l_{ws}$ segment, 
we set the value at $y_{start}-t$ to 1, and set all the other timestamps to 0. Second, if there is no OLV event performed in the sampled segment, we keep the zero-filled label list as the corresponding ground truths. 

However, the above approach to preparing for the labels is sub-optimal. The first reason is that, due to the burden of documenting clinical procedures, it happens that clinicians documented the OLV event a few minute later or earlier than the exact timestamp of the OLV procedural. In other words, the previous approach to generating the labels is vulnerable to label noise.
Second, 
optimizing the model using a label list that has a sudden high probability value at the exact timestamp and zeros nearby causes the model to generate discrete predictions. Then, it becomes more challenging to locate the OLV timestamps from the full-length signals. To tackle the weaknesses, we propose to utilize the Gaussian distribution function to smooth the ground truths. Specifically, we use a Gaussian distribution centered at the ground truth timestamp with the standard deviation $\sigma$, to construct the label distribution $y$ as shown in Equation \ref{eq:4}. 
\begin{equation}
{ y }_{ i }=\begin{cases} \frac { 4 }{ \sqrt { 2\pi  } \sigma  } exp\left( -\frac { { \left( { t }_{ i }-{ t }_{ OLV } \right)  }^{ 2 } }{ 2{ \sigma  }^{ 2 } }  \right) ,if-3\le{ t }_{ i }-{ t }_{ OLV } \le 3 \\ 0,\quad otherwise \end{cases}
\label{eq:4}
\end{equation}
Note that the smoothing step only applies to the $\pm3$ minutes near the ground-truth timestamp.  

As for the inference phase, similar to training, a sliding window with size $l_{ws}$, and a step size $l_{step}$ are used to split the signals into smaller equal-length segments. The total number of segments is calculated by:
\begin{equation}
n_{seg}=\operatorname{floor}\left(\frac{d-l_{ws}}{l_{step}}\right)
\label{eq_n_seg}
\end{equation}
where $d$ is the original sequence length. To obtain the final predictions for each patient, we find the maximum response among all the $n_{neg}$ outputs and the corresponding timestamp. Formally, 
\begin{equation}
    \hat { y } =argmax\left( \hat{y}_{seg}^{0};\hat{y}_{seg}^{1};...;\hat{y}_{seg}^{n_{seg}-1}  \right)
\end{equation}

\subsection{Loss function}
The model predicts the probability of an occurring OLV event for each timestamp. The binary cross entropy loss is used for optimization. In addition, we apply an auxiliary cross-entropy classification loss to the frequency feature to classify the binary occurrence of an OLV event.
\section{Results}\label{results}
In this section, we first describe our performance metrics and implementation details and then provide the comparisons with baseline methods. Following that, we analyze the importance of each variable through leaving-one-variable-out training and model interpretation. Lastly, we show the cross-institution generalization ability by training the model on Site A and testing on Site B, and vice versa.

\begin{table*}[h]
\centering
\begin{adjustbox}
{max size={\textwidth}{\textheight}}
 \begin{tabular}{c|c|c|c|c|c|c} 
 \hline
  \hline
 Method & MAE$\downarrow$ & acc5$\uparrow$ &acc4$\uparrow$ &acc3$\uparrow$ &acc2$\uparrow$ &acc1$\uparrow$\\ 
 \hline
  \hline
Change-point       & 15.3 $|$ 20.3 & 71.1 $|$ 73.1 & 68.5 $|$ 71.4 & 65.1 $|$ 67.6 & 55.8 $|$ 55.7 & 34.5 $|$ 31.5\\ 
Base-NN         & 19.0 $|$ 27.5 & 67.0 $|$ 62.8 & 63.9 $|$ 61.3 & 59.7 $|$ 58.0 & 53.3 $|$ 51.1 & 41.0 $|$ 36.8\\ 
Base-LSTM       & 6.0 $|$ 9.2 & 80.0 $|$ 84.3 & 77.1 $|$ 82.7 & 72.9 $|$ 80.4 & 66.7 $|$ 74.7 & 52.2 $|$ 57.8\\ 
w/o smoothing   & 5.8 $|$ 5.5 & 80.7 $|$ 89.1 & 78.1 $|$ 87.5 & 73.4 $|$ 85.4 & 67.1 $|$ 80.8 & 52.7 $|$ 67.1\\ 
w/o spectrum    & 4.5 $|$ 5.1 & 82.7 $|$ 89.9 & 79.6 $|$ 88.5 & 75.0 $|$ 86.1 & 68.6 $|$ 80.9 & 53.3 $|$ 64.5\\ 
Full model      & \textbf{4.4} $|$ \textbf{4.1} & \textbf{83.6} $|$ \textbf{91.3} & \textbf{80.2} $|$ \textbf{90.0} & \textbf{75.8} $|$ \textbf{87.6} & \textbf{69.4} $|$ \textbf{82.3} & \textbf{55.7} $|$ \textbf{67.2}\\
 \hline
  \hline
\end{tabular}
\end{adjustbox}
\caption{Model performances and ablation studies. \textbf{Change-point}: locating the maximum change point from the most OLV-correlated variable; \textbf{Base-NN}: baseline model using basic fully connected layers; \textbf{Base-LSTM}: baseline model by replacing transformer with LSTM cells. acc(n): the accuracy under a $n$ minutes margin, as discussed in the metrics section. The left and right side of "$|$" denotes OLV start and end, respectively.}
\label{table:1}
\end{table*}

\subsection{Model Performance}\label{metrics}
\textbf{Metrics.} Model performance is evaluated in the testing set. We use mean absolute error (MAE) and accuracy with error margin as the OLV timestamp estimation criteria. The justification of the event prediction accuracy is based on a pre-defined margin value, which is an allowable temporal range before or after the ground-truth timestamp. If the predicted timestamp is within this range, the prediction of the OLV event is a true positive. We use $acc3$ to represent accuracy with a clinically significant margin of 3 minutes. The MAE is calculated from the distance between the ground-truth timestamps and the predicted timestamps.

\begin{figure}[t!]
    \begin{minipage}{.5\textwidth}
    	\centering
    		\includegraphics[scale=0.21]{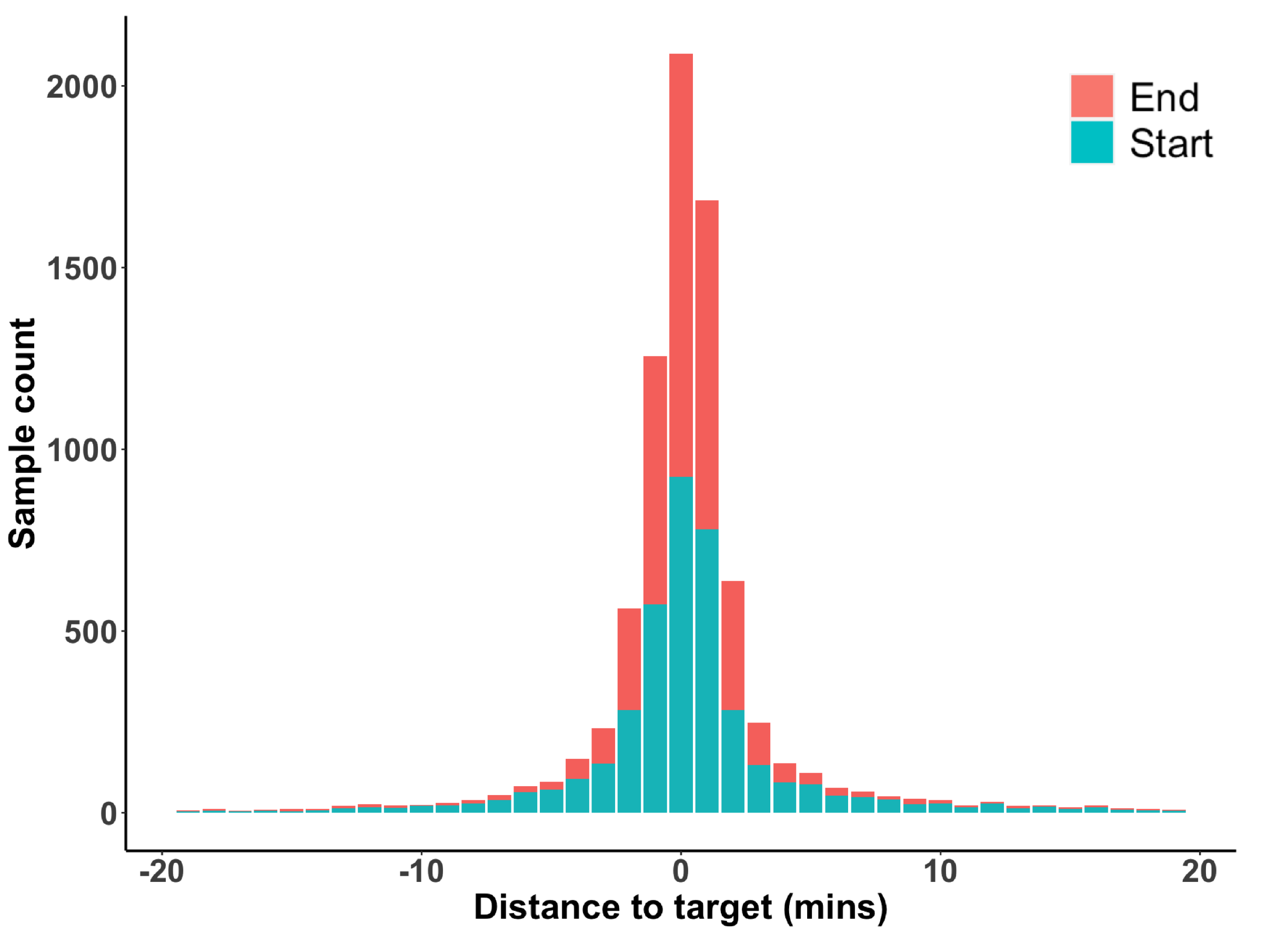}
    	  \captionof{figure}{Histogram of prediction errors \\for OLV start and end timestamps.}
       \label{fig5}
    \end{minipage}%
\begin{minipage}{.5\textwidth}
    \centering
    \includegraphics[scale=0.15]{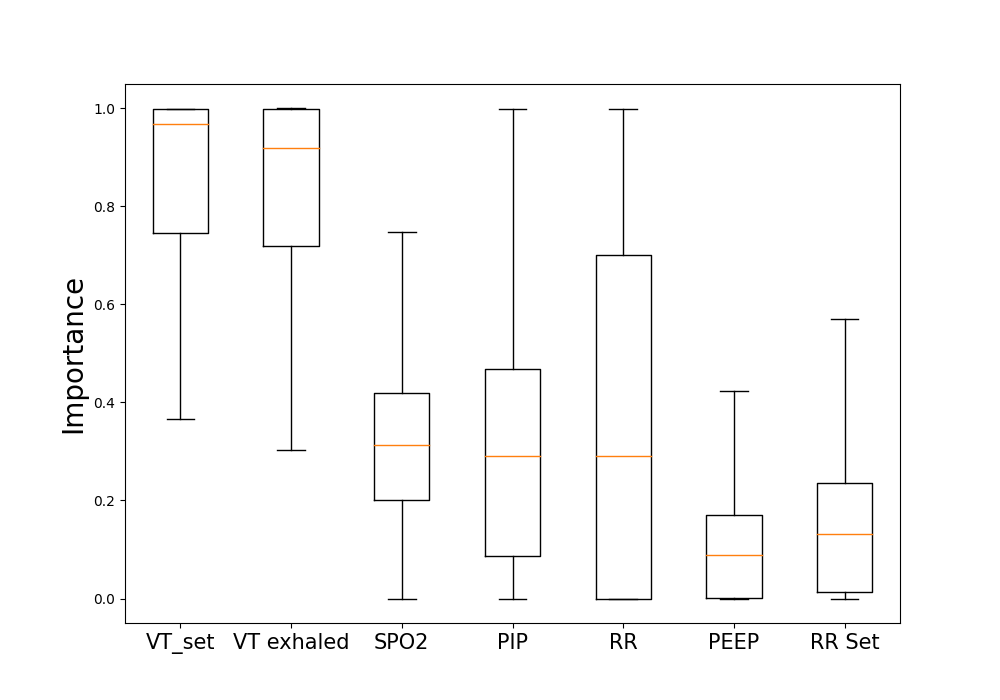}
    \captionof{figure}{Normalized attribution scores of each variable.}
    \label{fig:imp}
\end{minipage}
\end{figure}
\textbf{Implementation details}. Our framework is trained using PyTorch. The mini-batch size is 24, and the learning rate is set to 0.0005. We employed an Adam \cite{kingma2014adam} optimizer with 0.9 Nesterov momentum, where the weight decay rate is set to 0.0001. The sliding window size $ws$ is 40 minutes. We split the data into five folds randomly, and each fold includes 845 records. Five-fold cross-validation is applied for all experiments. The reported results are the average of every fold. Two identity models are trained separately to predict the respective start and end timestamps. All experiments are done in a GeForce RTX 2080Ti GPU. Detailed training and testing steps are shown in Algorithm \ref{alg:cap}.

\textbf{Overall performance and ablation study.} Table \ref{table:1} shows the performance using the full set of variables for training. Fig \ref{fig5} describes the histogram of the prediction errors for five folds of the testing set. The results show that most of the errors are centered around 0. A majority (81.7\%) of the predictions are within the 3 minutes error margin. The highest portion is at 0-minute error, where the model has successfully predicted the timestamp at the exact minute of the ground truth, and as the distance increases, the error percentage decreases significantly. We claim that most of the predictions are within a 3-minute error margin.

As compared to the baseline models, it can be seen from Table \ref{table:1} that our proposed consistently outperforms all the baselines by a considerable margin. We implemented a heuristic method that locates the top 2 change points from the most OLV-correlated variable (i.g., $VT\_set$). The results show that the hand-crafted feature based method produces inferior MAE and accuracy than data-driven learning-based methods. It's worth noting that the basic neural network with fully connected layers is not comparable to the sequential models such as LSTM (Long short-term memory) and Transformers. This indicates that incorporating temporal information is crucial for OLV detection. Although the LSTMs can learn temporal evolution using recurrent gates, they are sub-optimal compared to Transformers. Transformers enable direct connections between each node of timestamps, and the direct attention mechanism allows longer sequential modeling. 

In terms of the ablations as shown in Table \ref{table:1}, we observe an obvious decreasing in accuracy (from 83.6$|$91.3 to 80.7$|$84.3) and increasing of MAE (from 4.4$|$4.1 to 5.8$|$5.5) after removing the label temporal smoothing training. It demonstrates that smoothing the target timestamps during training produces score curves that have smoother transitions from OLV occur to not occur, consequently making it easier to locate the maximum response from the prediction curves. As shown in Fig \ref{fig4}, after smoothing the supervision signals with the Gaussian function, the predicted scores at each timestamp are also curved and noise reduced. On the contrary, without smoothing, the output scores present noticeable noise that interferes with locating the maximum score, thus yielding a biased predicted timestamp.

In order to see how much the frequency domain features affect the performance, we train our model again without the spectrum input and compare it to our original method. Utilizing the frequency domain features decreases the MAE from 4.5$|$5.1 to 4.4$|$4.1, and acc5 increases from 82.7$|$89.9 to 83.6$|$91.3. It shows the effectiveness of frequency features that can provide solid extra information for time-series OLV detection.

\subsubsection{Cross-site generalizability.} Recall that our dataset is curated from two institutions. Regarding the within-site results, we observe that site B yields better performance than site A. This makes intuitive sense because Site B consists of more patients' records than Site A. To investigate the model generalization ability across two sites, we use the data from one site for training and the data from the other for testing. Regarding cross-site validation, the MAE increases to 5.8$|$5.5 and 8.7$|$6.5. While underlying reasons remain unclear for further investigation, we attribute this to the distribution shift between the data from the two sites introduced by differences in patient demographics and clinical characteristics. We leave it as our future work to incorporate domain adaptation modules to fine-tune our models to improve the model generalizability \cite{farahani2021brief}.

\begin{table*}[h]
\centering
\begin{adjustbox}
{max size={\textwidth}{\textheight}}
 \begin{tabular}{c|c|c|c|c|c|c|c} 
 \hline
  \hline
 Train & Test & MAE$\downarrow$  & acc5$\uparrow$  & acc4$\uparrow$  &acc3$\uparrow$  &acc2$\uparrow$  & acc1$\uparrow$   \\ 
 \hline
  \hline
Site A & Site A    & 7.5 $|$ 5.9 & 67.4 $|$ 88.2 & 60.7 $|$ 86.5 & 52.5 $|$ 82.9 & 39.7 $|$ 77.7 & 26.9 $|$ 65.0\\ 
Site B & Site B    & 3.6 $|$ 3.9 & 87.8 $|$ 91.7 & 85.3 $|$ 90.4 & 81.9 $|$ 88.8 & 76.3 $|$ 84.5 & 61.0 $|$ 71.5\\ 
Site A & Site B    & 5.8 $|$ 5.5 & 83.4 $|$ 89.5 & 80.0 $|$ 88.0 & 72.9 $|$ 84.9 & 59.7 $|$ 75.8 & 38.4 $|$ 52.1\\ 
Site B & Site A    & 8.7 $|$ 6.5 & 67.0 $|$ 86.4 & 60.8 $|$ 83.6 & 52.4 $|$ 78.0 & 42.7 $|$ 66.2 & 29.3 $|$ 42.4\\ 
 \hline
  \hline
\end{tabular}
\end{adjustbox}
\caption{Cross-institution validation results. The left and right side of "$|$" denotes OLV start and end, respectively.}
\label{table:2}
\end{table*}

\begin{table*}[ht!]
\centering
\begin{adjustbox}
{max size={\textwidth}{\textheight}}
 \begin{tabular}{c|c|c|c|c|c|c} 
 \hline
  \hline
 Metrics & MAE$\downarrow$  & acc5$\uparrow$  & acc4$\uparrow$  &acc3$\uparrow$  &acc2$\uparrow$  & acc1$\uparrow$   \\ 
 \hline
  \hline
w/ all         & 4.4 $|$ 4.1 & 83.6 $|$ 91.3 & 80.2 $|$ 90.0 & 75.8 $|$ 87.6 & 69.4 $|$ 82.3 & 55.7 $|$ 67.2\\ 
w/o VT(m)   & 6.5 $|$ 5.8 & 78.2 $|$ 88.0 & 75.1 $|$ 86.5 & 70.6 $|$ 84.2 & 64.8 $|$ 78.3 & 50.4 $|$ 62.4\\ 
w/o VT(s)   & 5.6 $|$ 5.1 & 82.0 $|$ 90.0 & 78.5 $|$ 88.4 & 74.2 $|$ 86.1 & 66.6 $|$ 80.5 & 49.8 $|$ 65.2\\ 
w/o PIP (m) & 4.6 $|$ 5.2 & 82.2 $|$ 89.6 & 79.4 $|$ 88.2 & 74.7 $|$ 85.6 & 68.2 $|$ 79.1 & 53.3 $|$ 63.1\\ 
w/o PEEP (m)& 4.5 $|$ 4.5 & 83.0 $|$ 90.7 & 79.8 $|$ 89.2 & 75.3 $|$ 87.0 & 69.0 $|$ 81.7 & 53.6 $|$ 66.6\\ 
w/o RR (m)  & 4.3 $|$ 4.7 & 83.4 $|$ 90.4 & 80.4 $|$ 89.1 & 75.6 $|$ 86.6 & 69.1 $|$ 81.5 & 53.4 $|$ 65.8\\ 
w/o RR (s)  & 4.6 $|$ 4.7 & 83.2 $|$ 90.6 & 80.3 $|$ 89.3 & 76.0 $|$ 86.7 & 69.0 $|$ 80.8 & 53.3 $|$ 65.8\\ 
w/o SPO2 (m)& 4.4 $|$ 4.3 & 83.3 $|$ 90.8 & 80.3 $|$ 89.3 & 75.8 $|$ 87.0 & 69.8 $|$ 81.9 & 54.4 $|$ 66.5\\ 
 \hline
  \hline
\end{tabular}
\end{adjustbox}
\caption{Training by leaving one variable out. "s" and "m" means setting variables and measurement variables, respectively. Detailed variable descriptions are shown in Table \ref{tab:1}. The left and right side of "$|$" denotes OLV start and end, respectively.}
\label{table:4}
\end{table*}
\subsection{Feature Importance Decomposition}
To explore how important each variable contributes to the prediction model training and which variable has more information to predict the OLV start and end timestamp. We conduct our experiments in two aspects: (1) training without one of the variables and (2) computing the integral of the gradients of the output prediction for the predicted label with respect to the input variables \cite{sundararajan2017axiomatic}.

As can be seen from Table \ref{table:4}, after removing either of the $VT\_set$ or the $VT\_exhaled$, the MAE drops more significantly than the others. In contrast, the rest variables have a minor contribution to accuracy. Some variables contain easier-recognized signal patterns driven by OLV operation than others, which is consistent with our observations. By feeding all the variables to the model, the best performance is achieved. We can see that although some variables are much less informative than $VT\_set$ or $VT\_exhaled$, they still provide supplementary information for OLV prediction and contribute to higher accuracy. Figure \ref{fig:imp} shows the average of the attribution score of each variable of the patients using the integral of the gradients. Consistently, we observe higher attribution scores for $VT\_Set$ and $VT\_exhaled$. Following them, $PIP$ and $RR$ present a high importance score less frequently. Moreover, the scores are widely scattered between 0 to 1, which indicates that there exists a small portion of cases when $PIP$ and $RR$ contribute the most and act as dominant variables. The deep learning model adaptively attends to the variables that are correlated most to OLV events, and it lowers the reliance on the variables that hardly show OLV-correlated patterns. In contrast, the anomaly detection methods using hand-craft statistical patterns are incapable of weighing the reliance on the variables except by using pre-defined weights and thresholds.




\section{Discussion and Limitation}
In this paper, we developed and validated an innovative Transformer-based deep learning model for predictions of start/end timestamps of OLV procedures utilizing objective physiological monitoring data. We obtained a satisfactory predictive performance for this DL algorithm, and this framework can be potentially extended to applications of clinical auto-documentation of other OR clinical events or procedures in other medical settings such as Intensive Care Units (ICU) or Post Anesthesia Care Units (PACU). Additionally, our extended experiments allow us to track down the contributing factors of observed between-institution variance, providing insights for interventions to improve the quality of collected data.

Our study has limitations. First, this work utilizes a retrospective observational clinical dataset, which is inevitably exposed to selection bias when sampled from the target patient population. As the first pilot study of this program of research, we decided to first focus on clinical cases containing only one OLV event due to practical considerations of data quality control and conditioning, and the convenience of modeling. This choice was based on assumptions that the majority of thoracic cases would undergo only one OLV procedure, and that OLV timestamps of one-OLV cases would be documented more reliably compared to multiple-OLV cases. However, such choices might have ruled out cases with higher clinical complexity and more/shorter OLV procedures hence introducing selection biases into our study sample. It partially explains the discrepancy of predictive performance in cross-site generalizability analysis due to different clinical characteristics by selection. We plan to further include and adjust clinical confounding factors to overcome this issue in our next research. Second, deep learning models are known to be difficult to interpret and prone to overfit the observed data. We will address this limitation by introducing the game-theory-based SHAP \cite{NIPS2017_7062} values for explanations and minimizing over-fitting by utilizing internal cross-validation techniques and performing cross-sample (Site A vs. Site B) validations to assure model generalizability across samples from different institutions. We plan to refine our algorithm in followed studies to predict both non-event and multiple-events by incorporating external validation data sources from other institutions such as Columbia University Hospital systems. 


\bibliographystyle{spbasic}
\bibliography{spbasic}
\end{document}